\documentclass[
 reprint,
 superscriptaddress,
 amsmath,amssymb,
 aps,
pra,
showkeys
]{revtex4-2}

\usepackage{soul}
\usepackage{float}
\usepackage{multibib}
\usepackage{hyperref}
\usepackage{xcolor}
\hypersetup{
    colorlinks,
    linkcolor={blue!50!blue},
    citecolor={blue!50!blue},
    urlcolor={blue!80!black}
}
\usepackage{siunitx}
\usepackage{import}
\usepackage{placeins}
\usepackage{xcolor}
\usepackage{lmodern}
\usepackage[braket, qm]{qcircuit}
\usepackage{graphicx}
\usepackage{dcolumn}
\usepackage{bm}
\usepackage{comment}
\usepackage[english]{babel}
\usepackage{blindtext}
\usepackage{physics}  
\usepackage{multirow}
\usepackage{array}
\usepackage{enumerate}

\usepackage{ulem}

\begin{document}
\title{A Decoy-like Protocol for Quantum Key Distribution: Enhancing the Performance with Imperfect Single Photon Sources}  

\author{Chanaprom Cholsuk}
\email{chanaprom.cholsuk@tum.de}
\affiliation{Department of Computer Engineering, TUM School of Computation, Information and Technology, Technical University of Munich, 80333 Munich, Germany}
\affiliation{Munich Center for Quantum Science and Technology (MCQST), 80799 Munich, Germany}

\author{Furkan Ağlarc{\i}}
\affiliation{QLocked Technology Development Inc., 35430, Izmir, Türkiye}
\affiliation{Department of Physics, Izmir Institute of Technology, 35430 Izmir, Türkiye}

\author{Daniel K. L. Oi}
\affiliation{Computational Nonlinear and Quantum Optics, SUPA Department of Physics, University of Strathclyde, Glasgow G4 0NG, United Kingdom}

\author{Serkan Ateş}
\affiliation{QLocked Technology Development Inc., 35430, Izmir, Türkiye}
\affiliation{Department of Physics, Izmir Institute of Technology, 35430 Izmir, Türkiye}
\affiliation{Faculty of Engineering and Natural Sciences, Sabanci University, Tuzla, Istanbul 34956, Türkiye}

\author{Tobias Vogl}%
\email{tobias.vogl@tum.de}
\affiliation{Department of Computer Engineering, TUM School of Computation, Information and Technology, Technical University of Munich, 80333 Munich, Germany}
\affiliation{Munich Center for Quantum Science and Technology (MCQST), 80799 Munich, Germany}
\affiliation{Abbe Center of Photonics, Institute of Applied Physics, Friedrich Schiller University Jena, 07745 Jena, Germany}

\date{\today}

\begin{abstract}
Quantum key distribution (QKD) relies on single photon sources (SPSs), e.g. from solid-state systems, as flying qubits, where security strongly requires sub-Poissonian photon statistics with low second-order correlation values ($g^{(2)}(0)$). However, achieving such low $g^{(2)}(0)$ remains experimentally challenging. We therefore propose a decoy-like QKD protocol that relaxes this constraint while maintaining security. This enables the use of many SPSs with $g^{(2)}(0) > $0.1, routinely achieved in experiments but rarely considered viable for QKD. Monte Carlo simulations and our experiment from defects in hexagonal boron nitride show that, under linear loss, $g^{(2)}(0)$ remains constant, whereas photon-number-splitting (PNS) attacks introduce nonlinear effects that modify the measured $g^{(2)}(0)$ statistics. Exploiting this $g^{(2)}(0)$ variation as a diagnostic tool, our protocol detects PNS attacks analogously to decoy-state methods. Both single- and two-photon pulses consequently securely contribute to the secret key rate. Our protocol outperforms the Gottesman–Lo–Lütkenhaus–Preskill (GLLP) framework under high channel loss across various solid-state SPSs and is applicable to the satellite-based communication. Since $g^{(2)}(0)$ can be extracted from standard QKD experiments, no additional hardware is required. The relaxed $g^{(2)}(0)$ requirement simplifies the laser system for SPS generation. This establishes a practical route toward high-performance QKD without the need for ultra-pure SPSs.
\end{abstract}

\keywords{quantum key distribution, single photon source, second-order correlation function, GLLP}

\maketitle

\section{Introduction}
Quantum key distribution (QKD) has emerged as a promising technique for securing communications by harnessing the principles of quantum mechanics, in contrast to classical cryptography methods, whose security depends on mathematical complexity \cite{10.1103/RevModPhys.74.145,10.1103/PhysRevLett.67.661,10.1103/RevModPhys.74.145,10.1103/RevModPhys.81.1301,10.1038/nphoton.2007.22}. A variety of QKD protocols have been proposed and implemented, such as the well-known BB84 and B92 protocols \cite{10.1016/j.tcs.2014.05.025,10.1103/PhysRevLett.68.3121}. These protocols typically rely on single photon sources, where individual photons act as flying qubits and are prepared in randomly chosen quantum states and transmitted over a quantum channel. In an ideal scenario, each pulse contains exactly one photon, ensuring that any attempt at eavesdropping disturbs the quantum state and can be detected.\\
\indent In practice, however, most QKD implementations employ attenuated laser sources, commonly referred to as weak coherent pulses (WCPs), which follow Poissonian photon-number statistics. These sources are widely used because they are inexpensive, technologically simple, and can be operated at high repetition rates, whereas true single photon sources remain complex and less accessible. The drawback of WCPs is their non-ideal emission, which leads to a finite probability of generating multi-photon pulses and thereby opens the system to photon-number-splitting (PNS) attacks \cite{10.1103/PhysRevA.51.1863,10.1088/1367-2630/4/1/344}. In such an attack, an eavesdropper can nondestructively extract one photon from a multi-photon pulse while allowing the rest to reach the legitimate receiver, thus gaining information during the key sifting process without revealing their presence.\\
\indent To counter this vulnerability, decoy state protocols have been widely adopted \cite{10.1103/PhysRevLett.94.230504,10.1103/PhysRevA.75.012312,10.1103/PhysRevA.72.012326}. These schemes enhance security by varying the pulse intensities between signal and decoy states, enabling the legitimate parties to estimate the photon number distribution and detect PNS attacks through statistical analysis \cite{10.1103/PhysRevLett.91.057901}. Despite their practical effectiveness, decoy protocols still use WCPs, which makes them relatively easy to implement but not yet true single photon sources (SPSs). An alternative approach is to employ single photons from imperfect solid-state quantum emitters, whose emission exhibits sub-Poissonian statistics \cite{10.1002/qute.202200059,10.1002/qute.202300038,10.1088/1367-2630/16/2/023021,10.1093/nsr/nwaf147,10.1103/PhysRevApplied.23.054022,10.48550/arXiv.2506.15520}. When the second-order correlation at zero delay ($g^2(0)$) is sufficiently low, the probability of multiphoton emission is strongly suppressed, thereby satisfying the security requirements for QKD \cite{10.1103/PhysRevA.66.042315}. This makes solid-state SPSs particularly attractive candidates for non-decoy implementations.\\
\indent These SPSs include, e.g., quantum dots \cite{10.1038/s41534-023-00800-x,10.1088/1367-2630/14/8/083001}, color centers in diamond \cite{10.1088/1367-2630/16/2/023021,10.1103/PhysRevLett.89.187901,10.1103/PhysRevA.95.022338}, defects in GaN \cite{10.1093/nsr/nwaf147,10.1103/PhysRevApplied.23.054022}, and defects in two-dimensional materials such as hexagonal boron nitride (hBN) \cite{10.1002/qute.202300038,10.1002/qute.202200059,10.1063/5.0186767,10.48550/arXiv.2501.13902,10.1039/D5TC02805A}. Among these systems, the defects in hBN have attracted particular attention due to their unique properties, including room-temperature operation, high photon out-coupling efficiency \cite{10.1038/nnano.2015.242}, excellent brightness, reliable single photon purity \cite{10.1063/5.0147560}, integration with photonics components \cite{10.1088/1361-6463/aa7839,10.1002/adom.202002218}, indistinguishability \cite{10.1103/PhysRevApplied.19.L041003,10.1021/acsphotonics.9b00314}, temporal polarization dynamics \cite{10.1021/acsnano.3c08940}, extensive defect availability \cite{10.3390/nano12142427,10.1021/acs.jpcc.4c03404}, and suitability in quantum applications \cite{10.1002/qute.2023003433,10.1002/adom.202302760,10.1002/adom.202402508,10.1063/5.0188597}. \\
\indent Nevertheless, achieving ultra-low $g^2(0)$ values is experimentally challenging due to factors such as background emission, imperfect excitation schemes, and non-ideal emitter properties, which can introduce unwanted multi-photon events. Given these practical limitations, it is worth reconsidering whether strict suppression of multiphoton probabilities is always necessary. Some earlier studies have suggested that the requirement for low $g^2(0)$ can be alleviated by incorporating a non-polarizing beam splitter to monitor PNS attacks \cite{10.1088/1367-2630/11/11/113033}. However, the introduction of additional components can complicate the setup and introduce further losses. More recently, coincidence-based protocols have been proposed to enhance the secret key rate in WCP \cite{10.1002/qute.202400685}, but equivalent strategies for solid-state SPSs remain largely unexplored.\\
\indent Thus, this work proposes a decoy-like protocol that tolerates a higher presence of multiphoton events (moderate-to-high $g^2(0)$ values) while preserving security through active monitoring of system behavior without requiring additional experimental components. To model the photon statistics of an imperfect quantum emitter, in particular beyond the second-order, we experimentally characterized a defect in hBN. We found that while most system parameters are affected by channel loss and transmission imperfections, $g^{(2)}(0)$ remains fundamentally source-specific and invariant under linear loss. In contrast, photon-number-splitting (PNS) attacks induce nonlinear loss processes that alter $g^{(2)}(0)$, leaving a distinctive statistical signature. It is important to emphasize that simply monitoring the channel loss is not an option to detect a PNS attack: the eavesdropper could (theoretically) replace a lossy quantum channel with something more efficient to balance the loss introduced by the attack. In addition, in practical scenarios such as satellite or long-distance terrestrial links, link efficiency naturally fluctuates significantly. We therefore exploit $g^{(2)}(0)$ variation as an intrinsic indicator of PNS attacks, analogous to conventional decoy protocols, where differences in transmission between signal and decoy states reveal eavesdropping. This strategy enables secure inclusion of two-photon contributions in the secret key rate. \\
\indent We then benchmark our protocol against the established Gottesman–Lo–Lütkenhaus–Preskill (GLLP) formalism \cite{10.26421/QIC4.5-1}, using our hBN defect measurements as a case study, and extend the analysis to reassess key rates achievable with other solid-state SPS platforms. Crucially, since most QKD systems already employ a beam splitter and two single photon avalanche diodes (SPADs) with rare exceptions using electro-optic modulator (EOM)-based setups, $g^{(2)}(0)$ can be extracted directly from standard detection data. Our method is thus implemented entirely in post-processing, requiring no additional hardware. Overall, this work establishes a practical and broadly applicable framework for high-performance QKD that relaxes the conventional demand for ultra-low multiphoton emission and is compatible with any sub-Poissonian photon source.

\begin{figure*}[ht!]
    \centering
    \includegraphics[width=1\linewidth]{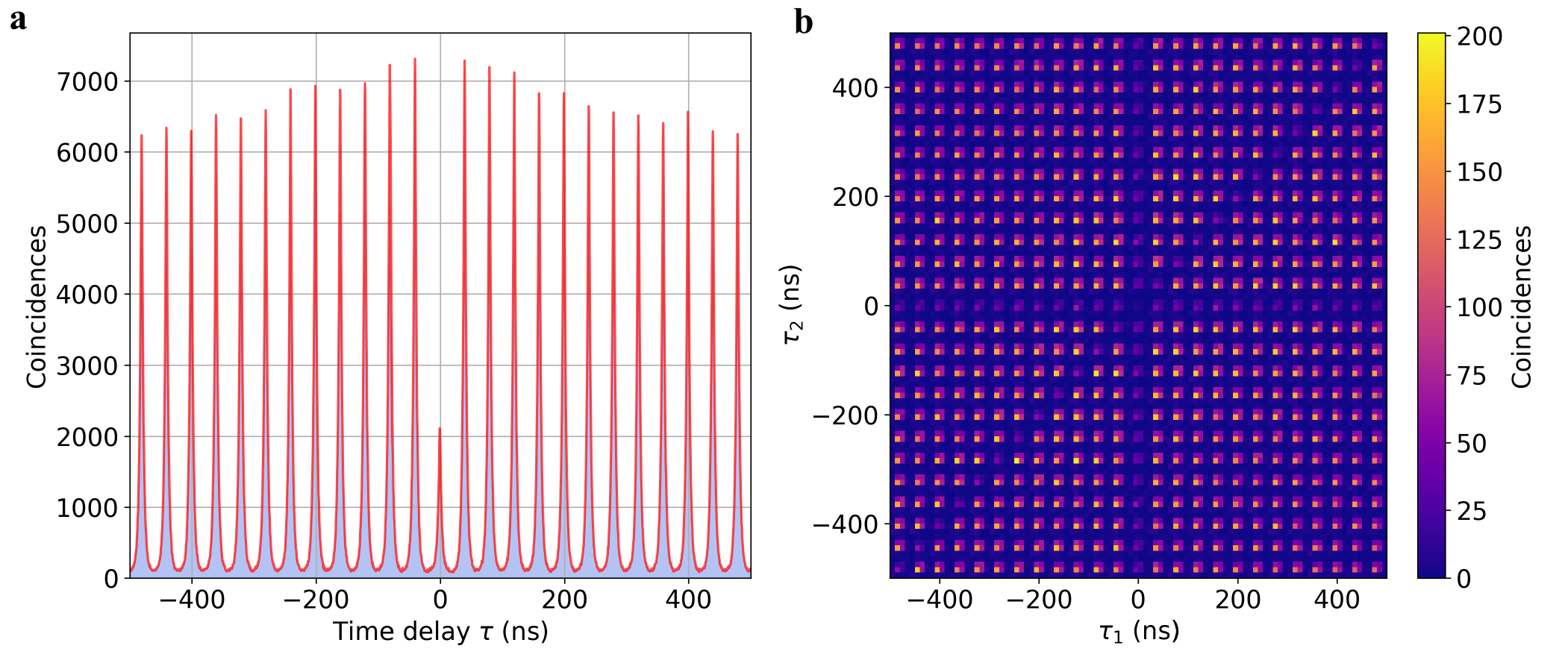}
    \caption{\textbf{Measured photon statistics of our hBN quantum emitter.} \textbf{a}
    Second-order ($g^{(2)}(\tau)$). \textbf{b} Third-order ($g^{(3)}(\tau_1,\tau_2)$) photon correlation functions under pulsed excitation at a 25 MHz repetition rate and 80.5 $\mu W$ excitation power. The delay range of $\pm20~\mu s$ is chosen in the calculation to ensure that the side peaks become flattened; however, here we show the zoomed $\pm 500~ns$ window for the sake of clarity.}
    \label{fig:g2_g3_hBN}
\end{figure*}

\section{Results}
This section presents and analyzes the performance of the proposed decoy-like protocol. We begin by deriving the corresponding secret key rate. Next, we characterize the photon statistics of our hBN defects. We then investigate how the photon number distribution and photon statistics evolve under varying attack strengths. Finally, we evaluate the resulting impact on the secret key rate and compare this decoy-like protocol with the conventional GLLP formalism.

\subsection{Proposed QKD formalism}
In this framework, the value of $g^{(2)}(0)$ is continuously monitored during operation as an indicator of PNS attacks, based on observable changes in photon statistics (to be demonstrated in the next subsection). As long as no statistically significant deviation in $g^{(2)}(0)$ is detected, it is assumed that there is no PNS attack, and both single-photon and two-photon pulses are allowed to contribute to the secret key rate, similarly to how keys are generated from signal and decoy states in decoy protocols \cite{10.1103/PhysRevA.72.012326,10.1103/PhysRevLett.94.230504}.\\
\indent To incorporate both contributions, we first consider the secret key rate $R$ in the asymptotic regime. The rate depends on the information Eve could potentially gain about each pulse, quantified by the mutual information $I(A;E)_n$. The secret key rate is then given by \cite{10.1002/qute.202400685}
\begin{equation}
    R \ge \frac{1}{2} \Big\{- Q_\mu h_2(E_\mu) f(E_\mu) + \sum_{n=1}^{\infty} Q_n (1 - I(A;E)_n) \Big\},
\end{equation}
where $Q_\mu$ is the total gain; $Q_n$ is the gain of the $n$-photon state; $h_2$ is the Shannon entropy function; $f(E_\mu)$ is the error correction efficiency; and $E_\mu$ is the quantum bit error rate (QBER).\\
\indent For pulses prepared in non-orthogonal states $\ket{\psi_0}$ and $\ket{\psi_1}$ with overlap $\cos c = \braket{\psi_0}{\psi_1}$, the overlap of the $n$-photon state is $(\cos c)^n$, which limits Eve’s distinguishability. Using the Holevo bound, the mutual information is bounded by \cite{10.1364/OE.25.011894}
\begin{equation}
    I(A; E)_n \le h_2\left(\frac{1 + (\cos c)^n}{2} \right).
\end{equation}
For multi-photon pulses ($n \ge 3$), which are much more vulnerable than single- or two-photon pulses, Eve can, in principle, exploit the additional photons to gain complete knowledge of Alice's bit without introducing errors. Specifically, she could perform a PNS attack, where one or more photons are diverted to her while the remaining photons reach Bob, or use unambiguous state discrimination (USD) to perfectly distinguish non-orthogonal quantum states \cite{10.1103/PhysRevA.62.022306}. To account for this worst-case scenario, we assume that Eve gain full information on all pulses with $n \ge 3$, that is, $I(A;E)_n = 1$.\\
\indent With these assumptions, the final expression for the key rate reduces to \cite{10.1364/OE.25.011894,10.1002/qute.202400685}
\begin{eqnarray}
    R \ge \frac{1}{2} \Big\{- Q_\mu h_2(E_\mu) f(E_\mu) + Q_1 (1 - \Phi(2 e_1 - 1)) \nonumber \\
    + Q_2 (1 - \Phi((2 e_2 - 1)^2)) \Big\}, \label{eq:SKR_monitoring}
\end{eqnarray}
where $\Phi(a) = h_2 \left( \frac{1}{2} + \frac{a}{2} \right)$ and $e_n$ is the error rate of the $n$-photon state. See Sec.~\ref{sec:keyrate} for further derivations.\\
\indent Eq.~\eqref{eq:SKR_monitoring} indicates that the privacy amplification gain receives contributions not only from single photon states but also from two-photon states. This additional contribution suggests the potential for achieving a higher secret key rate, as will be demonstrated in the following section.

\begin{figure*}
    \centering
    \includegraphics[width=1\linewidth]{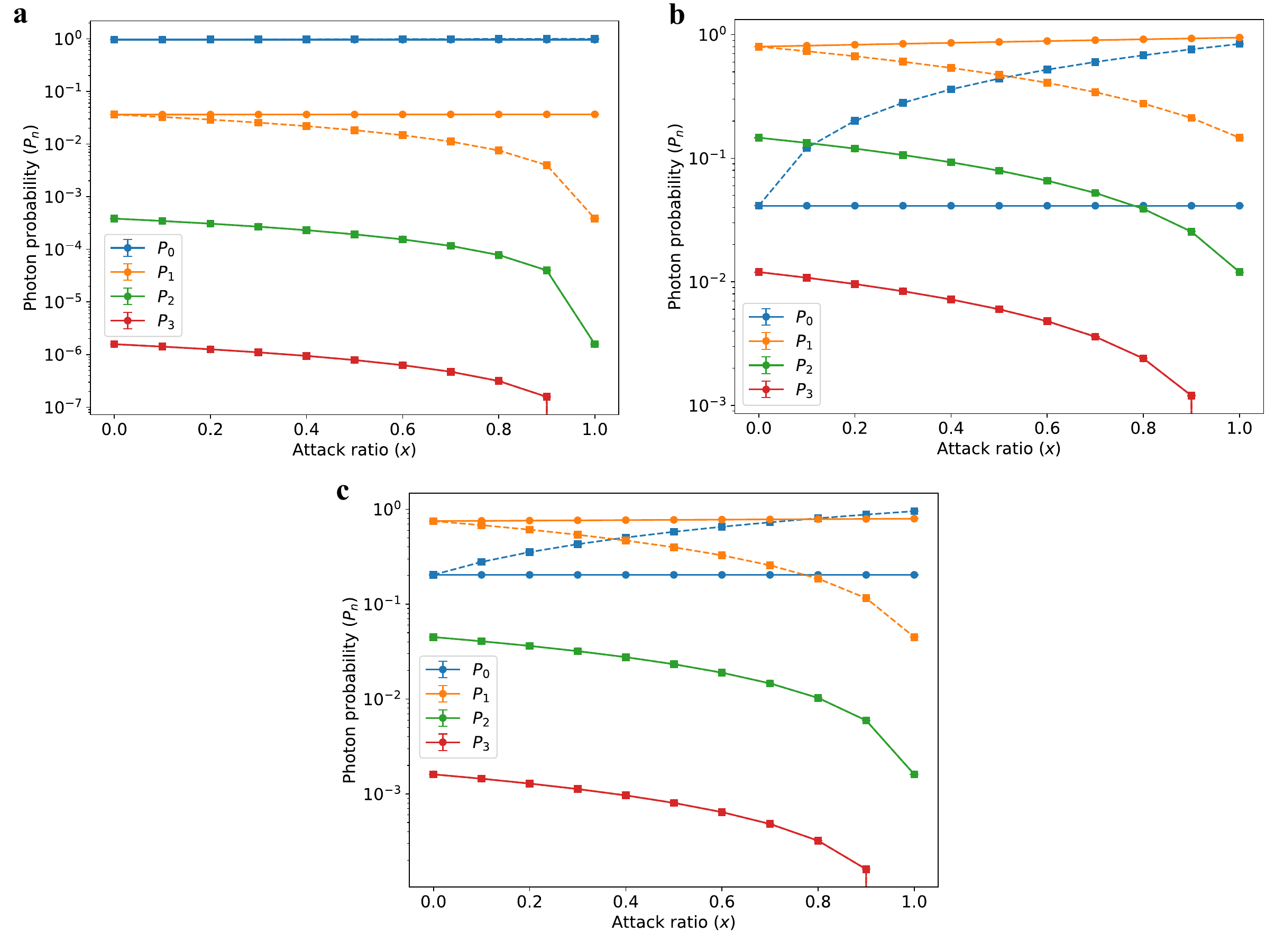}
    \caption{\textbf{Simulated photon number distributions $P_n$ after soft and hard PNS attacks, obtained via Monte Carlo simulation.} The attack strength $x$ is varied from 0 (no attack) to 1 (full PNS attack). \textbf{a} is for our hBN. \textbf{b} is for hBN with high quantum efficiency \cite{10.1364/OPTICA.6.001084}. \textbf{c} is for QD \cite{10.1103/PhysRevA.90.023846}. The solid and dashed lines represent the soft and hard PNS attacks, respectively. Each simulation was performed $100$ times with $10^7$ samples per run. The reported $P_n$ values represent the mean with standard deviations.}
    \label{fig:change_photonstat}
\end{figure*}

\subsection{Photon statistics of hBN quantum emitter}
To assess the impact of PNS attacks on photon statistics, it is essential to obtain the full distribution of photon number probabilities ($P_0$, $P_1$, $P_2$, $P_3$), corresponding to the vacuum, one-photon, two-photon, and three-photon states, respectively. While most prior studies report only the $g^{(2)}(0)$, and only a few extend to higher-order statistics, comprehensive measurements of $P_0$ to $P_3$ remain scarce. To address this gap, we experimentally characterized the photon statistics of our hBN quantum emitter under pulsed excitation. Fig.~\ref{fig:g2_g3_hBN} presents the measured photon correlation functions at an excitation power of 80.5~$\mu$W and a repetition rate of 25~MHz. Additional measurements performed at other excitation powers and repetition rates are provided in Supplementary Section~S1.  \\
\indent Fig.~\ref{fig:g2_g3_hBN}\textbf{a} shows the $g^{2}(\tau)$ histogram. Coincidence peaks within a delay range of $\pm 20~\mu$s are integrated with each peak summed over a 40~ns window, corresponding to the inverse of the pulse-repetition period ($1/f_{\text{rep}}$). This integration window ensures that the full contribution from every excitation pulse is included. In principle, one could apply the time filtering to reduce coincidences and then obtain a lower $g^{2}(0)$; however, this would lose some photons and lead to a low mean photon number. To allow consistent comparisons across measurements, we therefore fix the window width to $1/f_{\text{rep}}$ in all cases.\\
\indent Using this procedure, we obtain $g^{(2)}(0) = 0.559$, which is well below the classical (Poissonian) limit of $g^{(2)}(0)=1$ and only slightly above the single photon benchmark of 0.5 \cite{10.1088/1367-2630/ab3ae0}. This confirms the sub-Poissonian nature of the hBN emission, while indicating a modest residual multi-photon contribution under the present operating conditions. \\
\begin{figure*}
    \centering
    \includegraphics[width=1\linewidth]{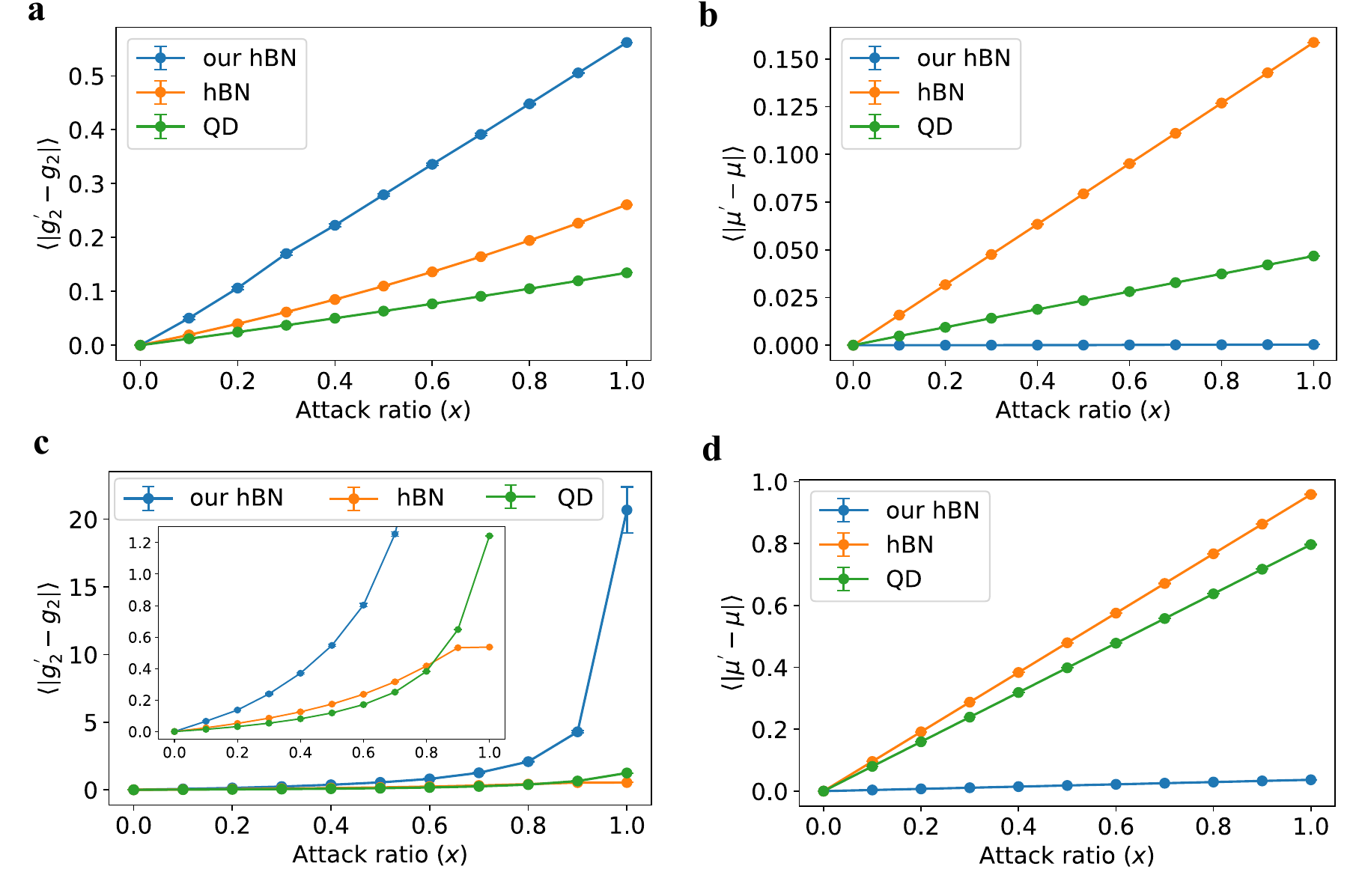}
    \caption{\textbf{Absolute changes in photon statistics of different physical systems under varying PNS attack ratios. }\textbf{a,c} the second-order correlation function at zero delay, $|g^{2}(0)' - g^{2}(0)|$, and \textbf{b,d} the mean photon number, $|\mu' - \mu|$, as functions of the PNS attack strength $x$. These results quantify how photon number splitting attacks progressively alter the photon statistics of the source under both soft and hard PNS attack models. \textbf{a,b} is for soft PNS attack while \textbf{c,d} is for hard PNS attack. Each simulation was performed $100$ times with $10^7$ samples per run. The reported $|g^{2}(0)' - g^{2}(0)|$ and $|\mu' - \mu|$ values represent the mean with standard deviations.}
    \label{fig:change_g2_mu}
\end{figure*}
\indent Fig.~\ref{fig:g2_g3_hBN}\textbf{b} displays the two-dimensional coincidence histogram used to evaluate $g^{(3)}(\tau_1,\tau_2)$. The lattice of coincidence peaks reflects the periodic pulsed excitation with a pronounced suppression at the central position corresponding to $g^{(3)}(0,0)$. By examining cross sections at $\tau_1=0$ and any $\tau_2$, or at any $\tau_1$ and $\tau_2=0$, the data reduce to effective second-order correlations, reproducing the $g^{(2)}(0)$ value. At the central point ($\tau_1=\tau_2=0$), we extract $g^{(3)}(0,0)=0.185$, providing direct evidence of higher-order antibunching. This result not only confirms the quantum nature of the hBN emission but also highlights the presence of genuine sub-Poissonian statistics beyond the second order. Comparisons with other hBN emitters and detailed fitting analyses are presented in Supplementary Section~S1.

\subsection{Effects of PNS attacks on photon statistics}
Next, we investigate how PNS attacks influence the photon statistics of different solid-state sub-Poissonian sources with varying quantum efficiencies. Specifically, we consider our hBN defects with a quantum efficiency of $\sim$3.63\%, hBN defects with a reported quantum efficiency of $\sim$80\% \cite{10.1364/OPTICA.6.001084}, and quantum dots (QDs) with a quantum efficiency of $\sim$75\% \cite{10.1103/PhysRevA.90.023846}, as shown in Figs.~\ref{fig:change_photonstat}\textbf{a}$-$\textbf{c}.\\
\indent In this work, two attack models are considered: soft and hard PNS attacks. In the soft attack, when a multi-photon pulse is emitted, Eve is assumed to capture exactly one photon and allow the remaining photons in the same pulse to reach Bob undisturbed. For example, if two photons are emitted, Eve keeps one and forwards one; if three are emitted, she keeps one and forwards two. As a result, some probability weight is shifted from higher-order photon terms ($P_2$, $P_3$, …) into the single photon term ($P_1$), as described by Eq.~\eqref{eq:soft_PNS} in the Methods section. In the hard attack, Eve is assumed to block all single photon pulses entirely, while applying the same strategy to multi-photon pulses as in the soft attack. This results in a redistribution of probabilities from multi-photon states into lower-photon-number states, while part of the single photon probability is additionally reassigned to the vacuum state, as expressed by Eq.~\eqref{eq:hard_PNS}.\\
\indent For Fig.~\ref{fig:change_photonstat}\textbf{a}, prior to the attack, the probabilities of $P_0$, $P_1$, $P_2$, and $P_3$ are 0.963, 0.036, 0.00038, and 1.57$\times$10$^{-6}$, respectively. Once the attack is applied, $P_0$ remains essentially unchanged, even under a full attack, for both soft and hard attack scenarios. This is because $P_0$ has a much higher ratio than the others. Even in the hard attack, where contributions from $P_1$ are reassigned to $P_0$, the change is negligible. In contrast, $P_1$ is unaffected by the soft attack but decreases under the hard attack, particularly as the attack approaches full strength. The probabilities $P_2$ and $P_3$ exhibit similar behavior in both soft and hard attacks, which is gradually decreasing with increasing attack proportion. For $P_3$ under a full soft or hard attack ($x=1$), the probability sharply drops to zero. This occurs because $P_3$ receives no contribution from $P_4$, as shown in Eqs.~\eqref{eq:soft_PNS} and \eqref{eq:hard_PNS} when $x = 0$ and $P_4 = 0$.\\
\indent For Fig.~\ref{fig:change_photonstat}\textbf{b}, before the attack, the probabilities are $P_0 = 0.041$, $P_1 = 0.800$, $P_2 = 0.147$, and $P_3 = 0.012$. Here, $P_0$ remains unchanged under the soft attack but increases under the hard attack, since part of $P_1$ is reassigned to $P_0$ according to Eq.\eqref{eq:hard_PNS}. Consequently, $P_1$ remains constant for the soft attack but decreases for the hard attack as its weight is transferred to $P_0$. The behavior of $P_2$ and $P_3$ is consistent with the previous case: both decrease monotonically with higher attack ratios, with $P_3$ eventually vanishing due to its already small contribution. For Fig.~\ref{fig:change_photonstat}\textbf{c}, before the attack, $P_0$, $P_1$, $P_2$, and $P_3$ are 0.203, 0.750, 0.045, and 0.002. In this case, the evolution of photon probabilities under both soft and hard attacks is similar to that observed in Fig.~\ref{fig:change_photonstat}\textbf{b}.\\
\indent Overall, these results show that the photon number probabilities evolve monotonically with attack strength. In particular, $P_1$, $P_2$, and $P_3$ decrease progressively as the attack ratio increases, while $P_0$ either remains nearly constant (soft attack) or increases slightly (hard attack), depending on the quantum efficiency of the source.\\
\indent We next examine how these redistributions in photon number probabilities affect the $g^{(2)}(0)$ and mean photon number ($\mu$), as depicted in Fig.~\ref{fig:change_g2_mu}. Before the attack, the $g^{(2)}(0)$ values are 0.559, 0.230, and 0.126 for our hBN, high-efficiency hBN, and QD sources, respectively, while the corresponding $\mu$ values are 0.037, 1.129, and 0.845. Under the soft attack, Figs.~\ref{fig:change_g2_mu}\textbf{a} and \textbf{b} indicate approximately linear variations in both $g^{(2)}(0)$ and $\mu$. Although the relative change is more pronounced for $g^{(2)}(0)$ than for $\mu$, both parameters follow the same trend: sources with larger initial values exhibit stronger variations under attack. Importantly, as already explained, monitoring $\mu$ is not a reliable strategy for detecting attacks. In practice, the losses are typically large, and then the eavesdropper could theoretically replace the quantum efficiency with something more efficient (such as a loss-less fiber) to balance the impact of the loss due to the attack. In contrast, $g^{(2)}(0)$ remains source-specific and insensitive to linear loss, making it a more robust parameter for attack detection.\\
\indent For the hard attack, Figs.~\ref{fig:change_g2_mu}\textbf{c} and \textbf{d} reveal a nonlinear dependence of $g^{(2)}(0)$ on the attack strength, while $\mu$ continues to alter linearly. Moreover, the overall changes under the hard attack are stronger than those observed under the soft attack. The same general trend can be applied; that is, sources with larger initial values undergo stronger variations, except in two cases. First, for our hBN at full hard attack ($x=1$), the absolute change in $g^{(2)}(0)$ increases sharply. This occurs because the initial $\mu$ is already small, and as $\mu$ approaches zero under this full attack, the calculated $g^{(2)}(0)$ diverges. Second, for high-efficiency hBN and QD sources, the relative trends of $g^{(2)}(0)$ interchange at large $x$. This behavior can also be explained by their mean photon numbers: since QDs start with a lower $\mu$ than high-efficiency hBNs, at high attack ratios, the further reduction in $\mu$ amplifies the change in $g^{(2)}(0)$ more strongly for QDs than for hBNs.\\
\indent In summary, the findings suggest that both soft and hard PNS attacks imprint clear signatures on $g^{(2)}(0)$. These systematic deviations can thus serve as reliable indicators of eavesdropping activity. Accordingly, real-time monitoring of $g^{(2)}(0)$ emerges as a practical countermeasure in QKD implementations, enabling the timely detection of potential PNS attacks.

\subsection{Secret key rate performance}
Having established that PNS attacks can be reliably monitored through changes in $g^{2}(0)$, it becomes possible to securely include contributions from two-photon pulses in the secret key rate calculation, as described in Eq.~\eqref{eq:SKR_monitoring}. In this section, we model how many photons must be received to estimate $g^{(2)}(0)$ with sufficient confidence. Then, we demonstrate the secret key rate performance and compare it with the GLLP formalism.\\
\begin{figure*}
    \centering
    \includegraphics[width=1\linewidth]{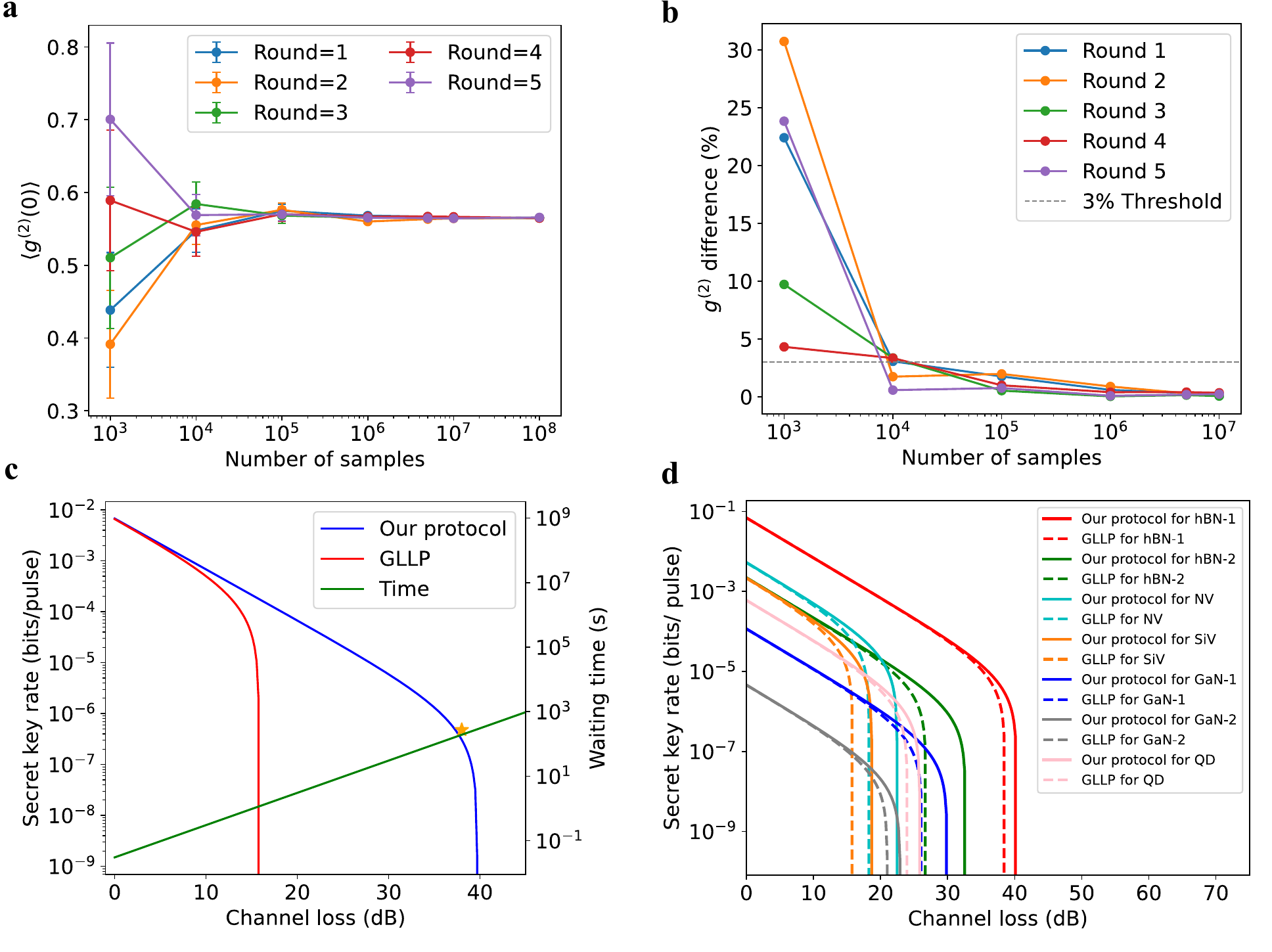}
    \caption{\textbf{Estimation of $g^{(2)}(0)$ and computed secret key rate as functions of channel loss for two scenarios.} \textbf{a} Convergence of $g^{(2)}(0)$ values across Monte Carlo samples for the hBN emitter. Each simulation round was repeated $100$ times. Reported $g^{(2)}(0)$ values represent the mean with standard deviations. \textbf{b} Differences in $g^{(2)}(0)$ across Monte Carlo samples for our hBN emitter, compared with the reference $g^{(2)}(0)$ obtained from $10^8$ samples. Each simulation round was repeated $100$ times. \textbf{c} Comparison of computed secret key rates between the conventional GLLP protocol and our proposed scheme for our hBN emitter. The secondary $y$-axis shows the waiting time required to receive $10^5$ photons at each channel loss. The orange star indicates the flyover duration and the 38-dB channel loss for the Micius satellite–ground link at 645 km \cite{10.1038/nature23655}. \textbf{d} Simulations for various solid-state quantum emitters using experimental parameters from Ref.~\cite{10.1002/qute.202200059} for hBN-1, Ref.~\cite{10.1002/qute.202300038} for hBN-2, Ref.~\cite{10.1088/1367-2630/16/2/023021} for NV and SiV in diamond, Ref.~\cite{10.1093/nsr/nwaf147} for GaN-1, Ref.~\cite{10.1103/PhysRevApplied.23.054022} for GaN-2, and Ref.~\cite{10.48550/arXiv.2506.15520} for QD. All QKD experiments are assumed to follow the BB84 protocol.}
    \label{fig:SKR}
\end{figure*}
\indent Fig.~\ref{fig:SKR}\textbf{a} presents the evolution of $g^{(2)}(0)$ as a function of the number of sent photons, which can be treated as the number of Monte Carlo samples. When only $10^{3}$ photons are considered, the mean $g^{(2)}(0)$ still exhibits noticeable fluctuations, even after averaging over 100 simulation runs. This indicates that this sample size is insufficient for stable estimation. As the number of photons increases, the $g^{(2)}(0)$ values gradually converge, with stability achieved after approximately $10^{5}$ photons. To quantify this convergence, we calculated the deviation of the simulated $g^{(2)}(0)$ from a reference value obtained using $10^{8}$ photons. For $10^{3}$ photons, the deviation remains large due to statistical uncertainty, whereas $10^{5}$ photons yield convergence within a 3\% confidence interval, as illustrated in Fig.~\ref{fig:SKR}\textbf{b}.\\
\indent Fig.~\ref{fig:SKR}\textbf{c} then compares the secret key rates obtained from our protocol and from GLLP for the hBN emitter, using the parameters summarized in Supplementary Section~S2. It suggests that the two schemes exhibit comparable performance up to roughly 10~dB of channel loss. Beyond this point, while both key rates decrease, the GLLP rate drops much more rapidly. This enhanced stability in our scheme arises from the inclusion of the two-photon pulse contribution, $Q_2$, in the key rate expression, as given in Eq.~\eqref{eq:SKR_monitoring}. Incorporating this term significantly extends the operational range and maintains the key rate at higher channel losses. In contrast, the conventional GLLP protocol treats multi-photon pulses as entirely insecure, assuming they are fully compromised by the PNS attacks. As a result, GLLP excludes the gain from two-photon and higher-order contributions, leading to a rapid decline in the key rate as channel loss increases. Our protocol can therefore enhance the secret key rate performance. Similar behavior is observed for different $g^{(2)}(0)$ values and excitation powers, as detailed in Supplementary Section~S3.\\
\indent We further estimate the time required to accumulate sufficient detection events under realistic transmission loss. As shown in Fig.~\ref{fig:SKR}\textbf{b}, at least $10^5$ photons must be sent to achieve a stable $g^{(2)}(0)$ value as explained earlier. Specifically, $10^5$ photons yield a deviation of less than 3\%. In practice, however, imperfect transmission increases the waiting time required to detect $10^5$ photons at the receiver. We therefore calculate the waiting time for each channel loss, as detailed in Sec.~\ref{sec:waiting_time}. We found that higher channel loss leads to longer waiting time, as depicted in Fig.~\ref{fig:SKR}\textbf{c}.\\
\indent To assess the practicality of our approach, we compare the required waiting times with the flyover durations of satellite QKD missions. For the Micius mission \cite{10.1038/nature23655}, which features a 645 km separation between the satellite and ground station and a link loss of approximately 38 dB, the reported flyover time is 273 seconds. Our results show that the required waiting time at this loss is shorter than the available flyover duration, implying that our protocol can support satellite-based QKD under these conditions. This result also underscores the robustness of our scheme: even a rudimentary hBN source, fabricated simply by drop-casting a flake and collecting it with an objective lens, already achieves compatibility with the Micius satellite link. For other missions, they are not included as they are either still planned or do not report the required parameters. Nevertheless, given typical satellite flyover times of 3–7 minutes, we expect our protocol to be applicable across a wide range of platforms.\\
\indent We further evaluated the secret key rate and waiting time for other SPSs. Remarkably, our protocol yields waiting times shorter than the Micius satellite flyover duration across all examined systems. Additional analysis is presented in Supplementary Section~S4.

\subsection{Re-evaluation of QKD experiments with single photon sources}
In this section, our proposed decoy-like protocol is applied to reevaluate the secret key rate of various solid-state SPSs from QKD experiments based on the assumption that they follow the BB84 protocol. Fig.~\ref{fig:SKR}\textbf{d} shows that, at low channel loss, the key rates obtained from our protocol and from GLLP largely overlap. However, as channel loss increases and the signal-to-noise ratio becomes limited by detector dark counts, the GLLP key rate for all sources rapidly falls to zero, whereas our protocol continues to provide a non-zero key rate due to the additional contribution from $Q_2$.\\
\indent Moreover, Fig.~\ref{fig:SKR}\textbf{d} indicates that some sources exhibit more significant improvements in the secret key rate than the others. For example, hBN-2, NV, and GaN-1 exhibit more pronounced enhancements compared with the remaining systems. We found that this improvement does not depend on individual source properties, such as $g^{(2)}(0)$. These findings suggest that our protocol can sustain high key-generation performance regardless of $g^{(2)}(0)$ quality, thereby relaxing the strict source-purity requirements imposed by conventional GLLP-based QKD.

\section{Discussion}
In this work, we have introduced a decoy-like QKD protocol that continuously monitors the $g^{(2)}(0)$ value to detect PNS attacks in real time. Combining Monte Carlo simulations with experimental photon statistics from our experiment in an hBN emitter, we observed that both $g^{(2)}(0)$ and $\mu$ are affected under a PNS attack. However, monitoring $\mu$ is unreliable, as high channel losses allow an eavesdropper to compensate the loss due to the attack through higher transmission efficiency. In contrast, $g^{(2)}(0)$ is intrinsic to the source and remains insensitive to linear loss, making it a robust and practical indicator for attack detection.\\
\indent By treating $g^{(2)}(0)$ as an operational security parameter, the protocol safely incorporates a contribution from two-photon pulses. This can enhance the secret key rate while relaxing the strict requirement for ultra-low $g^{(2)}(0)$ in single photon sources. This enables the use of many SPSs with $g^{(2)}(0)$ values above 0.1, routinely achieved in experiments but rarely considered suitable for QKD. Furthermore, this will also greatly simplify the excitation laser system. That is, in the GLLP protocol, the excitation laser pulse must be significantly shorter than the emitter’s excited-state lifetime to prevent multiple excitations within a single cycle, thereby ensuring near-ideal single-photon emission. This strict requirement necessitates the use of ultrafast picosecond or femtosecond laser systems. In contrast, our proposed decoy-like protocol can tolerate higher multi-photon pulses, allowing operation with higher $g^{(2)}(0)$. Consequently, the excitation pulse duration no longer needs to be much shorter than the excited-state lifetime; a pulse length comparable to the lifetime is sufficient. For hBN defects with a lifetime of approximately 3 ns \cite{10.1063/5.0147560}, a 1 ns excitation pulse can be appropriate. This relaxation of the timing constraint substantially simplifies the laser system requirements while maintaining secure key distribution.\\
\indent When applying our protocol to re-evaluate some existing solid-state quantum emitters, our results indicate that this protocol consistently outperforms the GLLP framework, demonstrating that high secret-key generation is achievable without imposing severe source-purity constraints. We also analyzed the waiting time required to accumulate sufficient detection events for stable $g^{(2)}(0)$ estimation of various SPSs. Our results indicate that, even under high channel loss, the required waiting time remains shorter than the available flyover durations of satellite QKD missions, such as Micius. This highlights the feasibility of deploying our protocol in realistic free-space and satellite-based scenarios.\\
\indent Importantly, the protocol relies solely on standard $g^{(2)}(0)$ measurements, which are accessible in nearly all solid-state SPS-based QKD setups. Therefore, it requires no additional experimental components. Last, this approach makes it possible to harness the large number of SPSs with moderate $g^{(2)}(0)$ values, well beyond just the ultra-low $g^{(2)}(0)$ sources traditionally considered for secure QKD. Overall, this work provides a practical and broadly applicable path toward high-performance QKD, enabling secure key distribution from many commonly available SPSs and paving the way for more accessible and scalable quantum communication systems.

\section{Methodology}
This section outlines the theoretical framework and simulation procedures used to evaluate the proposed QKD protocol under PNS attacks. The methodology consists of five parts. First, theoretical assumptions for modeling PNS attacks are established. Second, a Monte Carlo simulation is described to analyze the impact of the PNS attacks on photon number statistics. Third, the additional derivation of the secret key rate for the proposed protocol is presented. Forth, the summary of the GLLP formalism is provided. Last, the calculation detail for the waiting time analysis is explained.

\subsection{Theoretical assumptions for PNS attack}
The photon number statistics of a source can be described by the probability distribution of emitting $n$ photons per pulse, expressed as
\begin{equation}
P = 1 = \sum_{n=0}^{\infty} P_n = P_0 + P_1 + P_2 + P_3 + \cdots,
\end{equation}
where $P_n$ is the probability of emitting exactly $n$ photons in a given pulse. In particular, $P_0$ represents the probability of emitting a vacuum state (no photon), $P_1$ corresponds to a single photon emission, and $P_2$, $P_3$, and higher-order terms account for multi-photon emissions.\\
\indent For PNS attack, multi-photon events (i.e., pulses where $n \geq 2$) are of particular concern. An eavesdropper (Eve) could potentially perform a quantum nondemolition (QND) measurement \cite{10.1126/science.209.4456.547} to detect the presence of multiple photons without disturbing their quantum states. She could then split off one or more photons while forwarding the remaining photons to the legitimate receiver (Bob), thereby gaining information without introducing detectable errors into the quantum channel.\\
\indent To model this scenario, we consider two types of PNS attacks: soft and hard attacks. In a soft PNS attack, it is assumed that Eve splits off one photon from a multi-photon pulse, while leaving the remaining photons intact for Bob. This modifies the photon number distribution according to
\begin{eqnarray}
P_0' &=& P_0 \nonumber \\
P_1' &=& P_1 + x P_2 \nonumber \\
P_2' &=& (1-x) P_2 + x P_3 \nonumber \\
P_3' &=& (1-x) P_3 + x P_4 \nonumber \\
P_n' &=& (1-x) P_n \ \text{for}\ n \geq 4, \label{eq:soft_PNS}
\end{eqnarray}
where $P_n'$ represents the photon number probability after the attack, and $x$ is the attack ratio ranging from 0 to 1 (with $x = 1$ corresponding to a full soft attack applied to all multi-photon pulses).\\
\indent In contrast, a hard PNS attack assumes that Eve can completely block all single photon pulses ($P_1$) while selectively splitting multi-photon pulses as before. This results in the following photon number distribution after the attack
\begin{eqnarray}
P_0' &=& P_0 + x P_1 \nonumber \\
P_1' &=& (1-x) P_1 + x P_2 \nonumber \\
P_2' &=& (1-x) P_2 + x P_3 \nonumber \\
P_3' &=& (1-x) P_3 + x P_4 \nonumber \\
P_n' &=& (1-x) P_n \ \text{for}\ n \geq 4. \label{eq:hard_PNS}
\end{eqnarray}
Again, $x$ represents the fraction of pulses under attack, with $x = 1$ denoting a full hard attack.\\
\indent In this work, the photon number distribution is obtained from our experimentally characterized hBN defects (see experimental details in Supplementary S1). To apply this to the simulation, photon numbers up to $P_3$ are considered, as the probabilities for $P_{n \geq 4}$ are sufficiently small to be neglected.\\
\indent Here, $P_1$ corresponds to the quantum efficiency of the defect, while the multi-photon probabilities are given by
\begin{equation}
P_2 \leq \frac{1}{2} \mu^2 g^{(2)}(0), \qquad P_3 \leq \frac{1}{6} \mu^3 g^{(3)}(0,0),
\end{equation}
where $\mu$ is the mean photon number, and $g^{(2)}(0)$ and $g^{(3)}(0,0)$ are obtained from the experiment. The vacuum state probability is then given by
\begin{eqnarray}
    P_0 = 1 - P_1 - P_2 - P_3
\end{eqnarray}
Finally, $\mu$ is calculated from
\begin{equation}
\mu = \sum_{n=0}^{3} n P_n.
\end{equation}

\subsection{Monte-Carlo simulation for PNS attack}
To investigate the impact of PNS attacks on the photon number distribution and the associated photon correlation statistics, we performed a Monte Carlo simulation based on experimentally determined photon number probabilities. A sample of 10$^7$ photon emission events was generated to ensure statistical reliability, as increasing the sample size further does not significantly alter the results. Each simulation was repeated 100 times, and the reported values correspond to the mean over these runs.\\
\indent To simulate a PNS attack, we applied a probabilistic photon removal process, where each photon beyond the first in a multiphoton pulse is removed with a splitting ratio $x$ based on Eqs.~\eqref{eq:soft_PNS} and~\eqref{eq:hard_PNS}. After applying this transformation to the photon number distribution, both the original and attacked distributions were sampled, and the averaged values $\langle n \rangle$, $\langle n(n-1) \rangle$, and $\langle n(n-1)(n-2) \rangle$ were computed. These were then used to estimate the $g^{(2)}(0)$ and $g^{(3)}(0,0)$, respectively, according to
\begin{eqnarray}
    g^{(2)}(0) &=& \frac{\expval{n(n-1)}}{\expval{n}^2}, \\
    g^{(3)}(0,0) &=& \frac{\expval{n(n-1)(n-2)}}{\expval{n}^3}.
\end{eqnarray}
This process allows for evaluating how PNS attacks affect photon statistics and assessing the sensitivity of $g^{(2)}(0)$ as a real-time operational indicator for eavesdropping attempts.

\subsection{Secret key rate based on our protocol} \label{sec:keyrate}
To complement the secret key rate of our protocol proposed in Eq.~\eqref{eq:SKR_monitoring}, several additional parameters need to be computed. For the sake of completeness, we highlight some of the relevant parameters following Refs.~\cite{10.1063/5.0186767,10.1103/PhysRevA.72.012326} in this section. First, the gain needs to be computed. Here, the total gain ($Q_\mu$) can be determined from 
\begin{equation}
    Q_\mu = \sum_{n=0}^{\infty} Q_n,
\end{equation}
where the gain of the $n$-photon state ($Q_n$) can be derived from
\begin{equation}
    Q_n = Y_nP_n. \label{eq:Q1_ours}
\end{equation}
Here, $P_n$ represents the probability of n-photons and $Y_n$ represents the yield of $n$-photon states, which is
\begin{eqnarray}
    Y_n &=& Y_0 + \eta_\eta - Y_0\eta_\eta \approx Y_0 + \eta_\eta \nonumber \\
        &=& Y_0 + 1-(1-\eta_{tot})^n.
\end{eqnarray}
$E_\mu$ is QBER, which can be obtained from
\begin{equation}
    E_\mu Q_\mu = \sum_{n=0}^{\infty}e_nY_nP_n,
\end{equation}
where the error of the $n$-photon state ($e_n$) can be calculated from
\begin{equation}
    e_n = \frac{e_0 Y_0 + e_{int}\eta_n}{Y_n}.
\end{equation}
$f(E_\mu)$ represents the efficiency of error correction, which is widely assumed to be 1.22 \cite{10.1038/srep15276}. In the case where the detectors exhibit identical dark count rates, the baseline error $e_0$ is taken as 0.5, corresponding to uncorrelated background counts. The intrinsic system error $e_{int}$ is set to 3\%, a typical value observed in free-space QKD experiments \cite{10.1103/PhysRevApplied.18.024067,10.1038/nature23655}. Finally, The total efficiency, which accounts for channel loss, detector efficiency, and system misalignment, is given by
\begin{equation}
    \eta_{tot} = \tau_L \eta_{det},
\end{equation}
where $\tau_L$ is the channel transmission and $\eta_{det}$ is the detector efficiency.

\subsection{Secret key rate based on GLLP protocol} 
To access the performance of our proposed protocol, we compare the computed secret key rate with the GLLP based on the following expression  \cite{10.26421/QIC4.5-1,10.1103/PhysRevA.72.012326}
\begin{equation}
R_{\mathrm{GLLP}} = \frac{1}{2}\{ -Q_\mu h_2(E_\mu) f(E_\mu) + Q_\mu\Omega \left[ 1 - h_2\left( \frac{E_\mu}{\Omega} \right) \right] \}.
\end{equation}
$\Omega$ represents the fraction of untagged pulses, which in this work is estimated based on the measured $g^{(2)}(0)$ value as 
\begin{equation}
\Omega = 1 - \frac{P_2}{Q_\mu}.
\end{equation}

\subsection{Analysis of waiting time} \label{sec:waiting_time}
To evaluate how long it takes to accumulate the expected number of detected photons at a given channel loss, we first determine the number of photons required in the ideal case of perfect transmission. Using Monte Carlo simulations, we extract this photon number, which does not significantly alter the $g^{(2)}(0)$ compared with the highest sampling size. We then calculate the waiting time $T$ needed to receive the same number of photons under different loss conditions using
\begin{equation}
T = \frac{N}{f \times (1-e^{-\mu}) \times 10^{-\text{loss}/10} \times \eta_{\text{det}}}.
\end{equation}
Here, $N$ is the number of photons/samples (10$^5$ in this work), $f$ is the repetition rate (100 MHz in this work), $\mu$ is the mean photon number (0.037 for our hBN emitter), loss is the channel loss in dB, and $\eta_{\text{det}}$ is the detector efficiency (assumed to be 90\% based on typical SNSPD performance).

\section*{Data availability}
All raw data from this work is available from the authors upon reasonable request.

\section*{Notes}
The authors declare no competing financial interest.

\begin{acknowledgments}
This research is part of the Munich Quantum Valley, which is supported by the Bavarian state government with funds from the Hightech Agenda Bayern Plus. This work was funded by the Deutsche Forschungsgemeinschaft (DFG, German Research Foundation) under Germany's Excellence Strategy- EXC-2111-390814868 (MCQST) and as part of the CRC 1375 NOA project C2. The authors acknowledge support from the Federal Ministry of Research, Technology and Space (BMFTR) under grant number 13N16292 (ATOMIQS). SA acknowledges the support by the QuantERA II Programme under GA No 101017733 (Comphort), Scientific and Technological Research Council of Türkiye (TUBITAK) under GA Nos. 124N110 and 124N115.
\end{acknowledgments}

\section*{Author contributions}
T.V. conceived the project. C.C. carried out the simulations. F.A. and S.A. conducted the experiment. C.C, D.K.L.O., and T.V. analyzed the result. D.K.L.O. and T.V. supervised the project. All authors contributed to the review of the manuscript.

\bibliography{main}
\end{document}